\documentclass[%
 reprint,superscriptaddress,
 amsmath,amssymb,twocolumn
]{revtex4-2}

\usepackage{graphicx}
\usepackage{dcolumn}
\usepackage{bm}
\usepackage[utf8]{inputenc}
\usepackage{xcolor}
\usepackage{hyperref}
\usepackage{siunitx}
\usepackage{enumerate}
\usepackage{gensymb}
\usepackage{url}
\usepackage{xr}
\usepackage{textcomp}
\begin{document}

\title{Broadband Coherent Diffraction for Single-Shot Attosecond Imaging}

\author{Julius Huijts}
 \affiliation{LIDYL, CEA, CNRS, Université Paris-Saclay, CEA Saclay, Gif-sur-Yvette, France}
 
\author{Sara Fernandez}
\affiliation{LIDYL, CEA, CNRS, Université Paris-Saclay, CEA Saclay, Gif-sur-Yvette, France}

\author{David Gauthier}
\affiliation{LIDYL, CEA, CNRS, Université Paris-Saclay, CEA Saclay, Gif-sur-Yvette, France}

\author{Maria Kholodtsova}
\affiliation{LIDYL, CEA, CNRS, Université Paris-Saclay, CEA Saclay, Gif-sur-Yvette, France}

\author{Ahmed Maghraoui}
\affiliation{LIDYL, CEA, CNRS, Université Paris-Saclay, CEA Saclay, Gif-sur-Yvette, France}

\author{Kadda Medjoubi}
\affiliation{Synchrotron Soleil, BP 48, Saint Aubin, Gif-sur-Yvette, France}

\author{Andrea Somogyi}
\affiliation{Synchrotron Soleil, BP 48, Saint Aubin, Gif-sur-Yvette, France}

\author{Willem Boutu}
\affiliation{LIDYL, CEA, CNRS, Université Paris-Saclay, CEA Saclay, Gif-sur-Yvette, France}

\author{Hamed Merdji}\thanks{Corresponding Author: hamed.merdji@cea.fr}
\affiliation{LIDYL, CEA, CNRS, Université Paris-Saclay, CEA Saclay, Gif-sur-Yvette, France}

\date{\today}
%\graphicspath{{figures/}}

\begin{abstract}
Recent advances in the field of attosecond science hold the promise of tracking electronic processes at the shortest space and time scales. Imaging methods that combine attosecond temporal with nanometer spatial resolution are currently out of reach. Coherent diffractive imaging is based on the diffraction by a sample of a quasi-monochromatic illumination with a coherence time that exceeds the duration of an attosecond pulse. Due to the extremely broad nature of attosecond spectra, novel imaging techniques are required. Here, we present an approach that enables coherent diffractive imaging with a broadband isolated attosecond source. The method is based on a numerical monochromatisation of the broadband diffraction pattern by the regularised inversion of a matrix which depends only on the spectrum of the diffracted radiation. Experimental validations in the visible and hard X-rays show the applicability of the method. Because of its generality and ease of implementation for single attosecond pulses we expect this method to find widespread applications in future attosecond technologies such as petahertz electronics, attosecond nanomagnetism or attosecond energy transfer.
\end{abstract}

\maketitle

%%%%%%%%%%%%%%%%%%%%%%%%%%%%%%%%%%%%%%%%%%%%%%%%%%%%%%%%%%%%%%%%%%%%%%%%%%%%%%%%%%%%%%%%%%%%%%%%%%%%%%%%%%%%%%%%%%

\section{\label{sec:introduction} Introduction}
Extending coherent imaging, typically limited to quasi-monochromatic radiation, to broadband sources would open fascinating applications in attosecond science. Attosecond pulses which inherently exhibit broad spectra hold the promise to unravel Science and Nature's intricate machinery with ultimate details in both space and time. So far, attosecond science has been restricted to spectroscopy studies even though real space imaging has been envisioned  \cite{Krauz2009,Galmann2012,Callegari2016,Lindroth2019}. Table-top high-harmonic generation (HHG) sources can generate isolated attosecond pulses with a natural synchronisation with the driving laser field which makes them suitable for pump-probe studies at the shortest time scales. Isolated attosecond pulses in the microjoule range \cite{Takahashi2012}, thus compatible with single shot nanoscale imaging \cite{chapman2006a,Ravasio2009}, are available in the VUV. Recent advances are pushing HHG sources well into the soft X-ray regime, already covering the water window \cite{Teichmann2016} and up to the keV range \cite{Popmintchev2012} but with low photon output. Attosecond pulses at X-ray free electron lasers are also emerging \cite{Ding2009, Hartmann2018, Lindroth2019} promising orders of magnitude more peak power than table-top facilities.

Coherent Diffractive Imaging (CDI) is a powerful technique that has successfully validated single shot femtosecond nanoscale imaging \cite{chapman2006a,Ravasio2009} after its first experimental demonstration \cite{Miao1999}. By solving the so-called phase problem using a phase retrieval algorithm, CDI can be easily implemented. This lensless technique can reach spatial resolution limited only by the wavelength and with the possibility to extend to 3D perception \cite{Chapman2006, Nishino2009, Chapman2010, Ekeberg2015, Miao2015, Duarte2019}. However, CDI is based on the fundamental assumption of spatial and temporal coherence. Temporal coherence imposes the quasi-monochromaticity of the illumination. Its relevance to CDI can be explained from the Huygens–Fresnel principle: two points in a sample can no longer interfere if the difference in optical path length to the detector is larger than the coherence length $l_c = \pi c/\Delta\omega$, with $\Delta\omega$ the angular frequency bandwidth of the illumination and $c$ the speed of light. Therefore, the interference fringe visibility (the so-called speckle) decreases at the edges of the diffraction pattern, which restricts the useful numerical aperture of the diffraction pattern to low angles. This imposes a fundamental limit on the ultimate spatial resolution, $\Gamma$, given by $\Gamma=a \Delta\omega/\omega_c$, where $a$ is the largest dimension of the object being imaged and $\omega_c$ is the central angular frequency of the source. In practice, CDI can tolerate a finite bandwidth to get a successful phase retrieval. For example, soft X-ray femtosecond pulses from HHG and VUV FELs hold reasonable bandwidths, around 0.5 \%, and still allow CDI convergence \cite{chapman2006a,Ravasio2009}. However, performing CDI with current attosecond pulses poses a real challenge due to their extremely broad spectra exceeding 10 \% .  Novel strategies are required to solve the blurring of coherent diffractive patterns due to the broadband illumination and to allow for attosecond nanoscale imaging.

The first broadband phase-retrieval algorithm was introduced by Fienup to correct for aberration images from Hubble Telescope observations in the late 90s \cite{Fienup1999}. Recent works have explored multi-wavelength holographic imaging consistent with a train of attosecond pulses \cite{williams2015,Gonzalezthesis}, two-pulse coherent imaging \cite{Witte2014,Meng2015} or ptychography   \cite{Batey2014,Enders2014,Enders2016} (recently generalized by the group of Miao). Other successful methods are an extension of the commonly used Gerchberg-Saxton algorithm \cite{Gerchberg1972} in parallel to the iterative phase retrieval process dubbed PolyCDI  \cite{Abbey2011,Dilanian2009,Chen2009,Teichmann2010,Chen2012}, showing convergence of up to 3.7 \% bandwidth. In this work, we present a method compatible with single-shot attosecond flash imaging. We first present the principle of the method, then two experimental validations using visible and hard X-ray radiation.

%%%%%%%%%%%%%%%%%%%%%%%%%%%%%%%%%%%%%%%%%%%%%%%%%%%%%%%%%%%%%%%%%%%%%%%%%%%%%%%%%%%%%%%%%%%%%%%%%%%%%%%%%%%%%%%%%%

\section{\label{sec:theory} Numerical monochromatisation}

In a CDI experiment, the scattered light from a sample is detected in the far field. The diffraction pattern from the Fraunhofer diffracted field at a distance $z$ from the sample can be described as (see e.g. \cite{Paganin}):
\begin{equation}
\Phi_{\omega}(x,y,z) = \left(\frac{\omega}{cz}\right)^2 \left|{\psi}_{\omega}\left(k_x=\frac{\omega}{cz} x, k_y=\frac{\omega}{cz} y, z=0\right)\right|^2
\end{equation}
where $\omega$ is the angular frequency of the illumination and ${\psi}_{\omega}(k_x,k_y,z=0)$ denotes the 2-dimensional spatial Fourier transform of the field leaving the sample at $z=0$, with $k_x$ and $k_y$ the spatial frequencies. Notice that the far field distribution shows a scaling factor $\omega /cz$ that is wavelength-dependent.

\begin{figure}[t]
\includegraphics[width=0.5\textwidth]{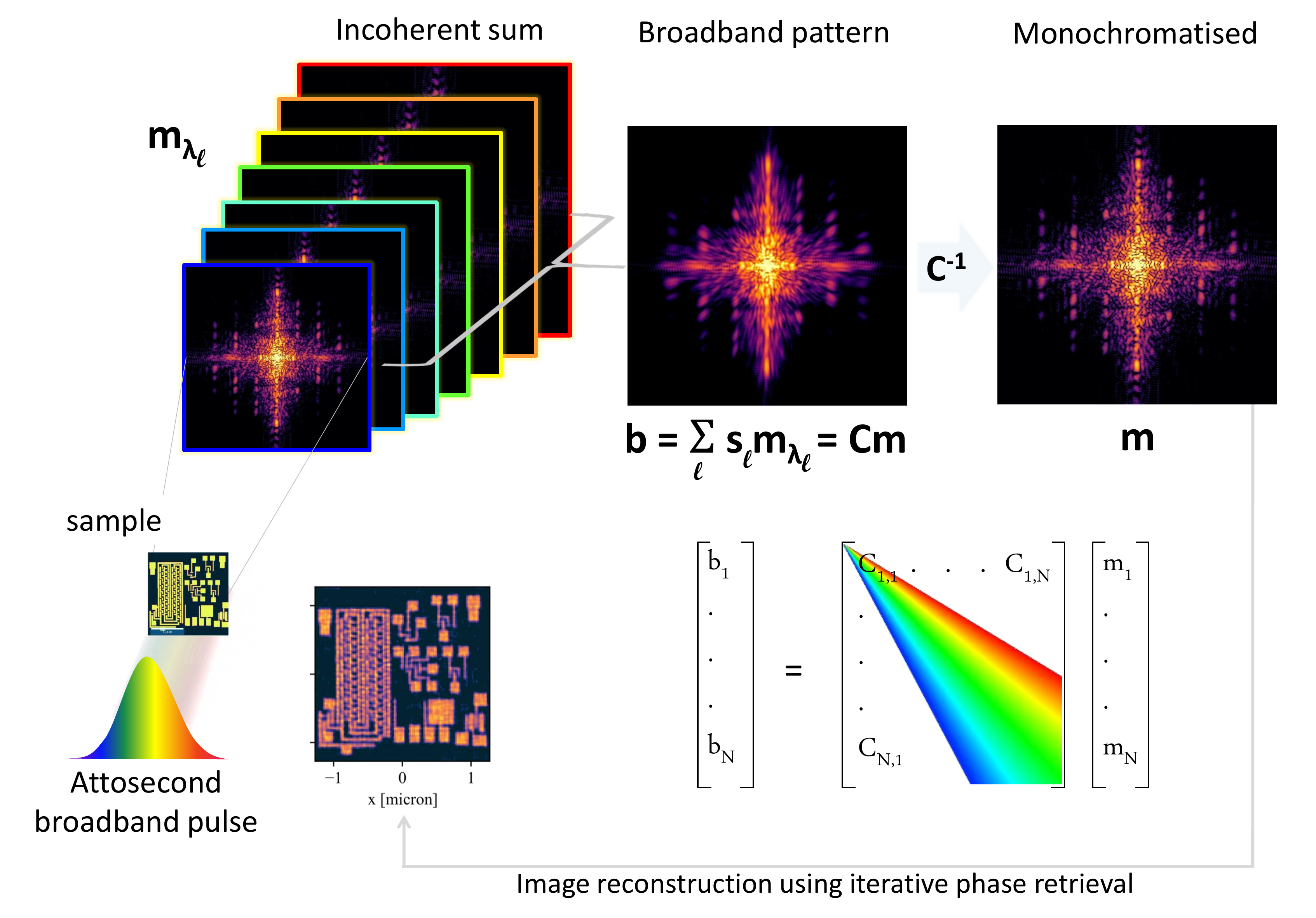}
\caption{\label{fig:broadband_CDI} \textbf{Principle of the numerical monochromatization.} Conventional CDI assumes a monochromatic source. In the case of an attosecond broadband source, the diffraction pattern is the incoherent, spectrally weighted, sum of the monochromatic diffraction patterns corresponding to all wavelengths present in the source. These monochromatic patterns are identical except for a geometric scaling. In the presented method this scaling is numerically inverted, thus yielding a monochromatised diffraction pattern. A conventional phase retrieval algorithm is then used to reconstruct the sample. The structure of the matrix $\mathbf{C}$ illustrates the method for the simplified case of a 1-dimensional diffraction pattern with the broadband and monochromatic patterns given by vectors $\mathbf{b}$ and $\mathbf{m}$ respectively. The monochromatisation method consists in the inversion of the matrix-vector problem $\mathbf{b}=\mathbf{Cm}$ in order to retrieve the monochromatic diffraction pattern from the broadband measurement.}
\end{figure}

We will now assume that ${\psi}_{\omega}(k_x,k_y,z=0) \approx s(\omega) {\psi}_{\omega_c}(k_x,k_y,z=0)$, with $s(\omega)$ the spectrum of the diffracted radiation through the sample ($s(\omega) \in\mathbb{C}$). We consider that the sample is spatially non-dispersive ($s$ does not depend on the sample coordinates) over a given spectral range around the central frequency $\omega_c$. This assumption is valid if the refractive indices of the materials in the sample do not change significantly over the spectrum of the source or if these changes are spatially homogeneous at the reconstructed length scales. Note that this assumption is also made in previous studies \cite{Abbey2011,Dilanian2009,Chen2009,Teichmann2010,Chen2012}. We can then express the diffraction pattern for a single angular frequency component:
\begin{equation}\label{eq:monoFraunhofer}
\Phi_{\omega}(x,y,z) = \omega^2 S(\omega) M_{\omega_c}\left(\frac{\omega}{cz} x,\frac{\omega}{cz} y\right),
\end{equation}
with $S(\omega)$ the power spectrum and $M_{\omega_c}$ the monochromatic diffraction pattern. Because the detector has an integration time much longer than the coherence time of the illumination, the broadband diffraction pattern is the incoherent sum of all angular frequency components, written as:
\begin{equation}\label{eq:convolution}
B(x,y,z) = \int \omega^2 S(\omega) M_{\omega_c}\left(\frac{\omega}{cz} x,\frac{\omega}{cz} y\right) \mathrm{d}\omega
\end{equation}
with the integral taken over the whole spectrum. The measured broadband diffraction pattern $B$ is thus given by the sum of scaled copies of the monochromatic diffraction pattern $M_{\omega_c}$, weighted by the normalised spectral density $S(\omega)$, such that $\int \omega^2 S(\omega) \mathrm{d}\omega = 1$. This is schematically shown in Fig. \ref{fig:broadband_CDI}.

In a conventional CDI experiment, the sample is reconstructed by the use of an iterative phase retrieval algorithm, which iterates between the monochromatic measured diffraction pattern and some constraints on the sample plane -typically the isolation of the object in real space- to converge to the solution. In our broadband case, a step of numerical monochromatisation is introduced prior to the iterative process. By measuring the broadband diffraction pattern $B$ and the spectrum of the diffracted radiation $S$, the retrieval of the monochromatic pattern $M_{\omega_c}$ is reduced to a linear algebra problem. Writing the monochromatic pattern as vector $\mathbf{m}$, the broadband pattern as vector $\mathbf{b}$ and the scaling matrix as $\mathbf{C}$, the broadband diffraction can be simply written as:
\begin{equation}
\mathbf{b} = \mathbf{C}\mathbf{m}.
\end{equation}
The matrix $\mathbf{C}$ can be regarded as containing the radially dependent point-spread function of the convolution in Eq. \ref{eq:convolution}. It maps a point in $\mathbf{m}$ to the shape of the spectrum in $\mathbf{b}$. This is schematically shown in Fig. \ref{fig:broadband_CDI}. For a 2-dimensional diffraction pattern $\mathbf{C}$ becomes a 4-dimensional tensor, although in the numerical implementation the diffraction patterns are rearranged in a 1-dimensional vector, so that $\mathbf{C}$ is kept 2-dimensional.

The monochromatisation of a broadband diffraction pattern is now reduced to the inversion of matrix $\mathbf{C}$. The matrix C is invertible as the determinant is non-zero. This is only the case for a spectrum that does not contain any zeros. Although $\mathbf{C}$ is invertible, the problem is ill-conditioned. This means that the inversion is very sensitive to noise, either experimental or coming from the discretisation. To mitigate this problem, a regularisation method known as Conjugate Gradient Least Squares (CGLS) \cite{Hansen1994} is used (See the Methods section and reference \cite{huijtsthesis} for a more detailed explanation of the numerical implementation). 

In the present method, Eq. \ref{eq:convolution} is thus deconvolved in a stand-alone step, that depends solely on the spectrum of the diffracted radiation. It can be used with standard phase retrieval algorithms, as shown in this paper, or integrated as a module in algorithms for other types of lensless imaging (holography, ptychography) to allow them to cope with broadband sources. Extensive simulations have been performed in \cite{huijtsthesis} on the method's applicability to ptychography.

%%%%%%%%%%%%%%%%%%%%%%%%%%%%%%%%%%%%%%%%%%%%%%%%%%%%%%%%%%%%%%%%%%%%%%%%%%%%%%%%%%%%%%%%%%%%%%%%%%%%%%%%%%%%%%%%%%

\section{\label{sec:exp}Experimental validation}
The monochromatisation method was first validated in the visible domain on a broadband and spatially coherent source, and then using synchrotron hard X-ray radiation. The samples employed for these experiments are ubiquitous in the field of diffractive imaging and allow for algorithmic validation and resolution estimation.

\subsection{\label{sec:visible}Visible light experiment}
\begin{figure*}[t]
\includegraphics{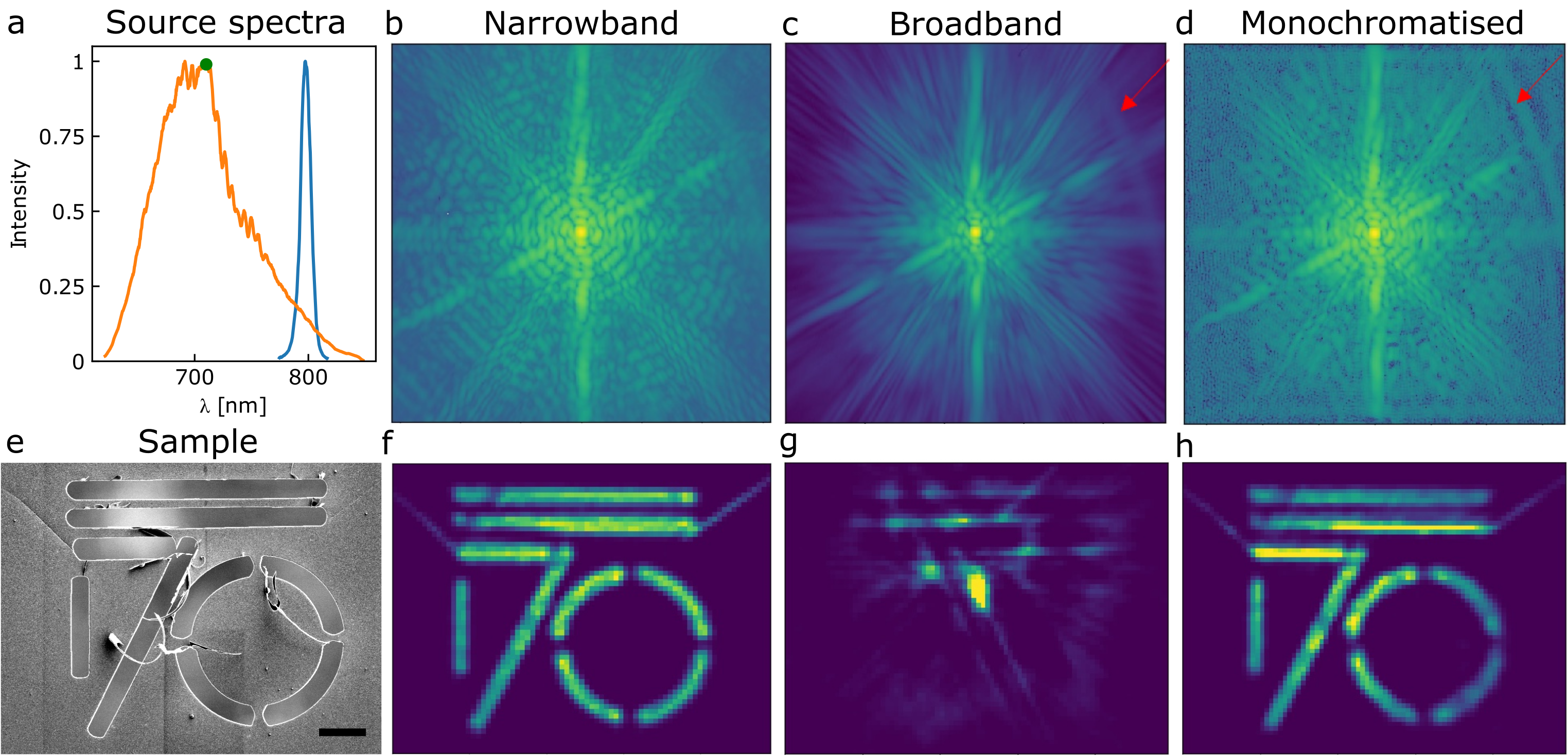}
\caption{\label{fig:results} 
\textbf{Experimental validation in the visible.}
 (a) The broadband spectrum (orange) with a bandwidth of $\Delta \lambda/\lambda=11\%$ and the narrowband spectrum (blue) with a bandwidth of $\Delta \lambda/\lambda=1.2\%$. The green dot at 710 nm represents the central wavelength $\lambda_c$, which is taken at the center of mass of the spectrum. (b) Narrowband diffraction pattern (c) Broadband diffraction pattern. Both patterns contain about $10^{11}$ photons. (d) Monochromatised diffraction pattern from the broadband data in (c). (e) SEM image of the sample. The scale bar corresponds to 10 \si{\micro\meter}. (f) Reconstruction for the narrowband case. (g) CDI reconstruction directly from the broadband pattern in (c). (h) Reconstruction of the monochromatised diffraction pattern in (d). Note how the cracks in the membrane are visible in both successful reconstructions.}
\end{figure*}

A scanning electron microscope image of the sample is shown in Fig. \ref{fig:results} (e). It consists of a gold-coated silicon nitride membrane in which features emulating the number ``70" were etched. The experiment was performed using laser pulses from an Yb:KGW femtosecond oscillator (1 W, 240 fs) focused into a photonic crystal fiber to generate a supercontinuum. After filtering the residual 1030 nm pump light, the resulting broadband spectrum with a bandwidth $\Delta\omega/\omega=\Delta \lambda/\lambda=11\%$ is shown in Fig. \ref{fig:results} (a). For the narrowband case ($\Delta \lambda/\lambda=1.2\%$), a 10 nm bandpass filter at 800 nm was used (Fig. \ref{fig:results} (a)). The light scattered from the object was recorded by a CCD camera set at 16 mm from the sample. Further details on the experimental setup can be found in Methods. When the sample is illuminated using narrowband radiation, the recorded diffraction patterns exhibit the expected coherent behaviour: size-modulated streaks arising from the upper lines of the ``70" and concentric rings corresponding to the ``0" shape dominate the pattern (see Fig. \ref{fig:results} (b)). By contrast, using the source’s full broadband spectrum yields a diffraction pattern that blurs the coherent modulations (see Fig. \ref{fig:results} (c)), as described in Eq. \ref{eq:convolution} and mentioned in Fig. \ref{fig:broadband_CDI}.

Our monochromatisation method was then applied to the broadband pattern, at the center of mass of the broadband spectrum, \textit{i.e.}, $\lambda_c=710$ nm. This allowed the successful recovery of the speckles and of most of the features. The similarity between the narrowband pattern and the monochromatised result (Fig. \ref{fig:results} (d)) is obvious. Some artifacts from the numerical process deserve nevertheless attention. It is the case, for instance, of the band of stray light indicated by a red arrow in the broadband pattern (Fig. \ref{fig:results} (c)). This stray light is not part of the real diffraction pattern and thus does not follow the spectral convolution expected by the inversion method. In the monochromatised pattern the inversion of this stray light causes unphysical oscillations (indicated by a red arrow in Fig. \ref{fig:results} (d)). Another unphysical feature that we systematically find is the squared structure at the edges of the monochromatised pattern. These regions are dominated by noise, which does not follow the expected spectral convolution either. Thanks to the regularised inversion method, these artifacts are not dominant in the monochromatised pattern and do not obstruct the phase retrieval.

The phase retrieval was performed using the diffraction patterns as input into a difference map algorithm \cite{Elser2007}, with a shrink-wrap-like support update \cite{Marchesini2007} (see Methods). The reconstruction procedure was applied to the three diffraction patterns displayed in Fig. \ref{fig:results} (b-d). The corresponding retrieved objects are shown in the lower row of the same figure. Whereas the CDI algorithm does not reach convergence when using the broadband diffraction pattern (Fig. \ref{fig:results} (g)) and only some lines are barely retrieved, the sample is readily reconstructed for both the monochromatic (Fig. \ref{fig:results} (f)) and numerically monochromatised cases (Fig. \ref{fig:results} (h)). In the monochromatised reconstruction, most of features visible in the SEM image are reproduced with high fidelity, even the cracks in the membrane to the left of the ``7" and to the right of the horizontal slits. However, the residual sub-micrometer gold wires visible from the patterning are not visible as they scale below the spatial resolution. A more inhomogeneous distribution of the intensity in the monochromatic case is observed, but the high quality of the result validates our approach.

\begin{figure}[t]
\includegraphics[width=0.5\textwidth]{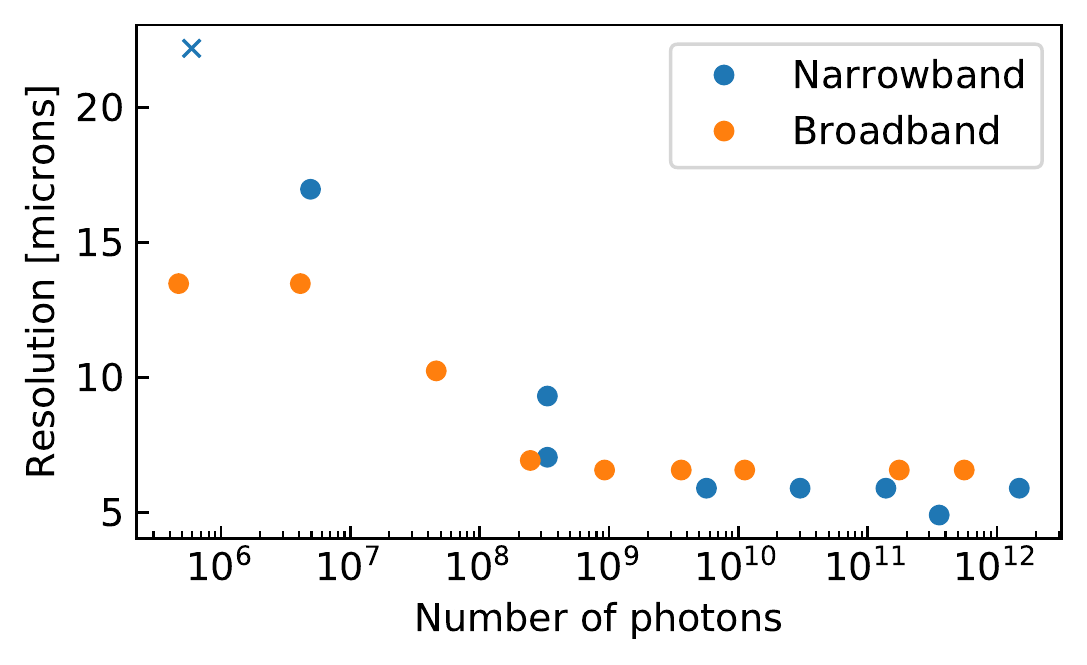}
\caption{\label{fig:res_sig} 
\textbf{Resolution of the reconstructions as a function of the signal level.}
 The resolution in the reconstruction, obtained from the 1/e-criterion on the PRTF is plotted as a function of the total number of photons in the raw measured diffraction pattern. Our broadband method performs similar to the conventional narrowband case and even better in the low signal regime. The cross indicates no successful reconstruction was achieved.}
\end{figure}

To test the method's sensitivity to noise, the experimental validation was reproduced for signal levels spanning six orders of magnitude. The resolution derived from the $1/e$-criterion on the phase retrieval transfer function (PRTF)\cite {Chapman2006} is plotted for each signal level in Fig. \ref{fig:res_sig}. Our broadband method performs similar to the narrowband case in the high-signal regime and slightly outperforms the narroband case in the low-signal regime. A possible explanation for this behaviour could be the noise-suppressing properties of the CGLS regularisation method used in the inversion of matrix $\mathbf{\mathrm{C}}$ (see e.g. \cite{Hansen1994}).
From Figure \ref{fig:res_sig}, it appears that the obtained resolution is flat over a signal increase of 2 orders of magnitude. We can give another estimate of the resolution by looking at the smallest feature that has been reconstructed reliably, which are the two oblique cracks in the membrane. They are visible in all high-signal reconstructions and disappear only below a signal level of $10^7$ photons. A lineout across these cracks in the reconstruction has a FWHM of about 2.5 microns. This performs better than the theoretical resolution limit of 5.5 \si{\micro\meter} determined by the coherence length for the general broadband case.

\subsection{\label{sec:xray}X-ray experiment}
In order to test our method in the hard X-ray regime, we performed a validation experiment at the SOLEIL synchrotron (France) at the Nanoscopium undulator beamline \cite{Somogyi2015}. Synchrotron radiation does not offer a broad spectrum of spatially coherent radiation, so that a broadband source was emulated by summing monochromatic diffraction patterns at different energies of the X-ray beam.

\begin{figure}[t]
\includegraphics[width=0.5\textwidth]{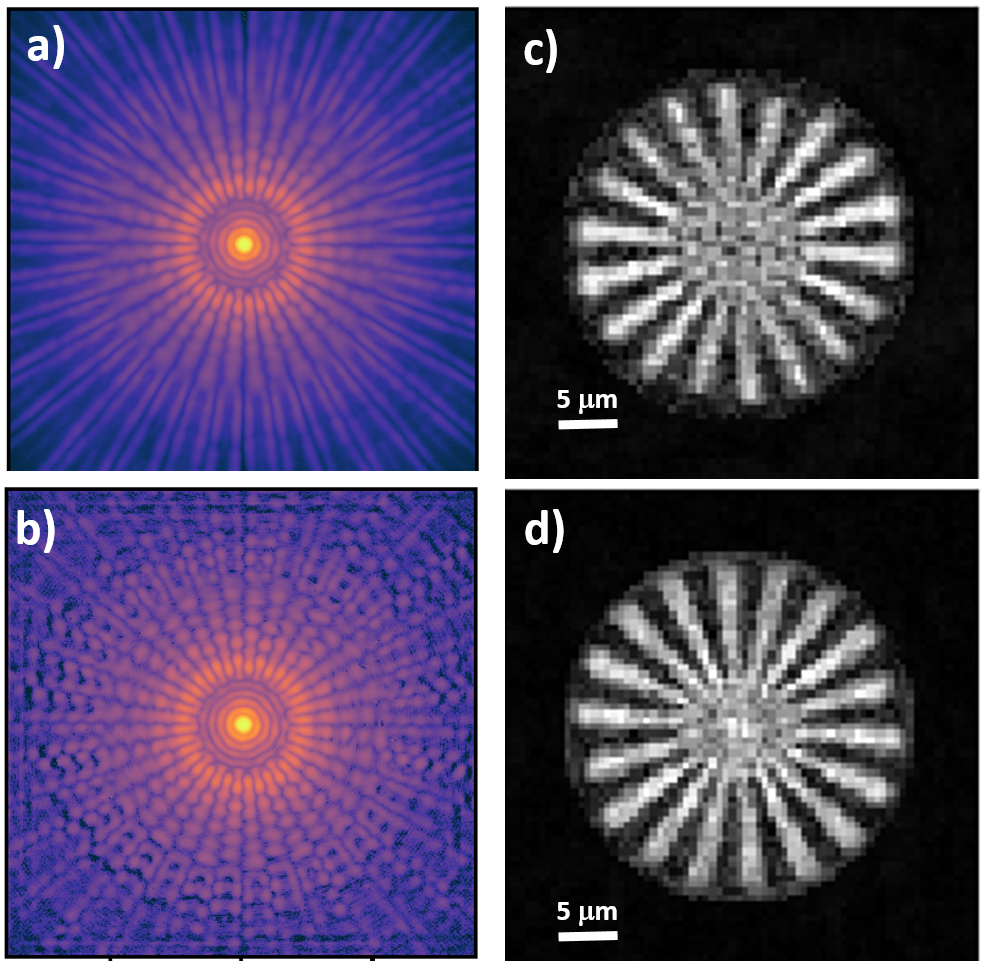}
\caption{\label{fig:soleil_x4} 
\textbf{X-ray experimental validation.}
(a) Broadband diffraction pattern obtained by summing all the acquired monochromatic diffraction patterns. (b) Diffraction pattern obtained after numerical monochromatisation of (a). (c) Reconstructed amplitude of the Siemens star sample from the monochromatised broadband diffraction pattern (b). (d) Reconstructed  amplitude of the same object from a single monochromatic diffraction pattern.}
\end{figure}

The sample chosen for this experiment consisted in a Siemens star resolution target with 10 \si{\micro\meter} length spokes and 20 \si{\micro\meter} total diameter (see methods). The sample was illuminated in transmission geometry to collect single diffraction patterns at successive energies, ranging from 7.1 to 8.0 keV in steps of 4 eV. The sum of the patterns yields a broadband diffraction pattern of 12\% bandwidth. As shown in Figure \ref{fig:soleil_x4} (a), the pattern shares some characteristics already observed in the experiment in the visible: it is blurred and it does not present the features that we would expect for a coherently illuminated object, such as the frequency modulation of the streaks arising from the star spokes. The application of our monochromatisation process was then performed on the pattern. The result is shown in Figure \ref{fig:soleil_x4} (b), now exhibiting sharp features and interference fringes. The appearance of artifacts near the borders is noticeable.

Phase retrieval was then performed using the \textit{PyNX} \cite{Mandula2016, pynxsite} package (see Methods). Figure \ref{fig:soleil_x4} (c) shows the reconstructed magnitude of the sample from the broadband pattern after monochromatisation. For comparison, the same phase retrieval procedure was applied to a single monochromatic diffraction pattern (at 7.1 keV), shown in figure \ref{fig:soleil_x4} (d). It should be noted here that the reconstructed pixel size is 560 nm, owing to the experimental configuration so that some details from the test sample are missing. However, this proof of principle experiment did not aim at resolving internal submicron features of the Siemen Star. Overall, the quality and the resolution of the broadband reconstruction is lower than that of the monochromatic benchmark which is attributed to instabilities when acquiring successively the monochromatic diffraction patterns.
%%%%%%%%%%%%%%%%%%%%%%%%%%%%%%%%%%%%%%%%%%%%%%%%%%%%%%%%%%%%%%%%%%%%%%%%%%%%%%%%%%%%%%%%%%%%%%%%%%%%%%%%%%%%%%%%%%
\section{Conclusions}
We have demonstrated successful image reconstruction from a single broadband coherent diffraction pattern paving the way towards single-shot attosecond lensless imaging. Our method is based on the numerical monochromatisation of the broadband diffraction pattern through the regularised inversion of the matrix-vector problem. It is a stand-alone step before application of the phase retrieval algorithm and depends only on the spectrum of the diffracted radiation. The method has been tested to be robust to low signal levels which can be pushed further using sparsity algorithm. We have successfully validated our method experimentally with more than 10\% bandwidth and up to the hard X-ray photon energy range. This applicability can be expanded to several tens of percent bandwidth (see supplementary information). The method can be applied to any technique based on coherent elastic scattering of any type of coherent radiation. Furthermore, since our technique applies as a prior step to the phase retrieval process itself, both single-shot CDI and ptychography could benefit from it.
The possibility of using broadband illumination in lensless imaging demonstrated here will open up a wide range of opportunities in attosecond science. Attosecond nanoscale flash imaging is now potentially accessible at a number of HHG facilities.  Attosecond XFELs and other broadband X-ray sources such as Inverse Compton Scattering sources \cite{Variola2014,Gunther2018} are being built and will benefit from our method. The 2-steps aspect (monochromatisation followed by phase retrieval) of the method allows a fast flow of the data which can benefit for example in real-time functional imaging in the semiconductor industry (see \cite{Zhang2013,huijtsthesis,IMEC2019} and supplementary information). The method being quite general, it can also be applied to various problem in modern optics such as aberration correction.

%%%%%%%%%%%%%%%%%%%%%%%%%%%%%%%%%%%%%%%%%%%%%%%%%%%%%%%%%%%%%%%%%%%%%%%%%%%%%%%%%%%%%%%%%%%%%%%%%%%%%%%%%%%%%%%%%%

\bibliography{broadband}

\begin{acknowledgments}
We acknowledge financial support from the European Union through the Future and Emerging Technologies (FET) Open H2020: VOXEL (grant 665207) and PETACom (grant 829153)  and the integrated initiative of European laser research infrastructure (LASERLAB-EUROPE) (grant agreement no. 654148). Support from the French ministry of research through the 2013 Agence Nationale de Recherche (ANR) grants ”NanoImagine”, 2014 ”ultrafast lensless Imaging with Plasmonic Enhanced Xuv generation (IPEX)”, 2016 “High rEpetition rate Laser for Lensless Imaging in the Xuv (HELLIX)”; from the DGA RAPID grant “SWIM”, from the Centre National de Compétences en Nanosciences (C’NANO) research program through the NanoscopiX grant; the LABoratoire d’EXcelence Physique Atoms Lumière Matière - LABEX PALM (ANR-10-LABX-0039-PALM), through the grants ”Plasmon-X” and “HIgh repetition rate Laser hArmonics in Crystals (HILAC)” and, finally, the Action de Soutien à la Technologie et à la Recherche en Essonne (ASTRE) program through the “NanoLight” grant are also acknowledged. 
The authors would like to acknowledge support of Franck Fortuna and Laurent Delbsq from CSNSM, IN2P3, Orsay for sample fabrication. We aknowledge Marc Hanna (LCF, IOGS Palaiseau), Florent Guichard (Amplitude Techologies), Michele Natile (Amplitude Techologies), Yoann Zaouter (Amplitude Techologies) for support during the experimental validation of the method in the visible. We aknowledge Guillaume Dovillaire and Samuel Bucourt (Imagine Optic, Orsay, France) for providing the CCD camera detector. We have also appreciated fruitful discussions with T. Auguste, F. Maia, H. Chapman, Liping Shi, Milutin Kovacev and B. Daurer on the principle and implementation of the method and acknowledge access to the Davinci computer cluster of the Laboratory of Molecular Biophysics (Uppsala University, Sweden) and support by M. Hantke on the use of \textit{Condor}. Contributions to the detector development  of Kewin Desjardins from SOLEIL (Saint Aubin, France) were crucial to the success of the synchrotron experiment. 
\end{acknowledgments}

\section*{Author contributions}
JH and HM proposed the physical concept. JH developed the monochromatisation method and performed the experiment in the visible, HM and JH devised the experiments, all authors performed the synchrotron experiment. SF and DG contributed equally. Monochromatisation of the synchrotron data was performed by JH, phase retrieval by SF. All authors discussed the results and contributed to writing the manuscript.

\section*{Competing interests}
The authors declare no competing interests

\appendix*

\section{Methods}
\subsection{Numerical implementation}
\begin{figure*}[t]
\includegraphics{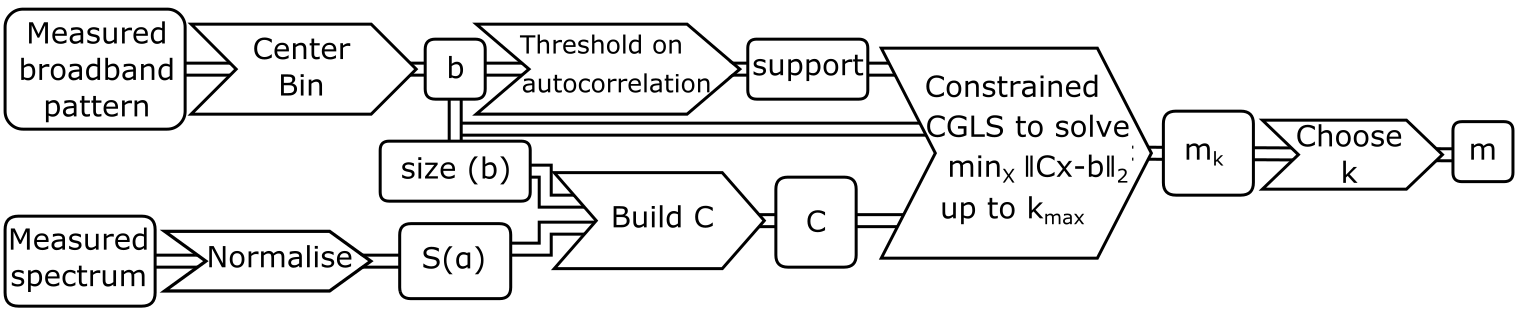}
\caption{\label{fig:flowchart} The information flow of our method.}
\end{figure*}

Matrix $\mathbf{C}$ is fully determined by the spectrum and the size of the diffraction pattern. Defining the scaling factor $\alpha=\lambda/\lambda_c$, in one dimension $\mathbf{C}$ is formed as follows:
\begin{equation}\label{eq:buildC}
C_{nj} = \sum_{L} \underbrace{\left[\min \{ j\, ,\, \alpha_ln \} -\max \{j-1\, ,\, \alpha_l(n-1) \} \right]}_\text{part of scaled pixel $n$ falling onto pixel $j$} \frac{S_l }{\alpha_l}
\end{equation}
where
\begin{align*}
\begin{split}
N&=\left\{n:\frac{j-1}{\alpha_{max}} < n < \frac{j}{\alpha_{min}+1}\right\},\\
L&=\left\{l: \frac{j-1}{n} < \alpha_l < \frac{j}{n-1} \right\}.
\end{split}
\end{align*}
Here $l,n,j$ are the indices that run over $\alpha, \mathbf{m}, \mathbf{b}$ respectively. This expression can be understood as: ``The contribution of pixel $n$ of $\mathbf{m}$ to pixel $j$ of $\mathbf{b}$ is given by the part of pixel $n$ that falls onto pixel $j$ for the scaled pattern $l$ times the spectral weight, summed over all $L$. ''
The flow of the numerical implementation of our method is depicted in Fig. \ref{fig:flowchart}.
In the experiment the broadband pattern and the spectrum are measured (this is either a corrected source spectrum or the spectrum of the diffracted radiation, and both should be corrected for the response of the camera). The measured pattern is centered and binned to obtain pattern $\mathbf{\mathrm{B}}$. The spectrum is normalised to obtain $S(\alpha)$ which, combined with the size of $\mathbf{\mathrm{B}}$, is all the information needed to compute matrix $\mathbf{\mathrm{C}}$. In our experimental validation in the visible, $\mathbf{\mathrm{B}}$ was 480x480 pixels. This means that $\mathbf{\mathrm{C}}$ has $240^4$ values (the scaling is the same for each quadrant). Luckily, $\mathbf{\mathrm{C}}$ is a sparse matrix, depending on the spectrum only a few percent of the values is non-zero. $\mathbf{\mathrm{C}}$ is now built by evaluating Eq. \ref{eq:buildC} in a set of parallel for-loops running only over the non-zero part of $\mathbf{\mathrm{C}}$. Note that $\mathbf{\mathrm{C}}$ only has to be built once per spectrum (and size of $\mathbf{\mathrm{B}}$). As $\mathbf{\mathrm{C}}$ is highly ill-conditioned, inversion of the problem is performed using a regularisation method called Conjugate Gradient Least Squares (CGLS, see e.g. \cite{Hansen1994}). It consists of minimising the least squares problem
\begin{equation}
\underset{x}{\mathrm{min}} \| Cx-b\|_2 \qquad \text{subject to} \quad x \in \mathcal{K}_k
\end{equation}
where $\mathcal{K}$ denotes the so-called Krylov subspace:
\begin{equation}
\mathcal{K}_k \equiv \mathrm{span}\{C^Tb,C^TCC^Tb, \dots,  (C^TC)^{k-1}C^Tb \}.
\end{equation}

The power of this method is its behavior of semi-convergence: for increasing $k$, first the signal is inverted so $x$ comes close to the exact solution, then the noise starts being inverted as well and x diverges (see \cite{Hansen1994,huijtsthesis} and the movies in the Supplementary Information). Typically the inverted patterns $M'_k$ are computed up to $k_{max}=40$, $M'_k$ is then visually inspected and the optimum value for $k$ is chosen (typically $k_{opt} \approx 25$, depending on the signal to noise ratio). The monochromatised pattern $M'_{k_{opt}}$ now serves as input for a conventional phase retrieval algorithm.
The numerical implementation is based on a Matlab function \cite{Hansen1994} which was translated to Python with the additions of two constraints: positivity of $x_k$ (photon counts should not be negative) and a support constraint on the Fourier transform of $x_k$. The latter is justified as we are dealing with isolated samples (the typical constraint for CDI), so the sample’s autocorrelation is isolated as well. These constraints help to further improve the regularising power of the method. 
The developed Python code is available on github.com/jhuijts under the BSD license.

\subsection{Experiments and data analysis}
\subsubsection{Optical experiment}
An off-axis parabola with an effective focal length of 75 mm was used to focus the beam on the sample, which was mounted on a translation stage. The sample consisted of a gold-coated silicon nitride membrane in which an aperture was etched using a focused ion beam. A CCD (Illunis RMV 4022) was used to acquire the diffraction pattern, at a distance of 16 mm from the sample. This corresponds to a numerical aperture of 0.39 for the diffraction patterns in Fig. \ref{fig:results}. These patterns were obtained by using high-dynamic range acquisitions combining acquisitions with different neutral density filters, such that the effective bit-depth of the shown images is up to 24. They subsequently underwent three pre-processing steps of a 3 x 3 binning yielding patterns of 480 x 480 pixels, numerical monochromatisation (broadband case only) and erosion outside a user selected mask (noise suppression). The sample was then reconstructed using a difference map algorithm with a shrink-wrap-like support, within a maximum of 1000 iterations. 1024 independent reconstructions were launched with random starting points to compute the PRTF. The successful reconstructions were then user-selected to be registered (using an adaptation from \cite{Guizar2008}) and averaged to enhance the signal on the final result.

\subsubsection{Synchrotron experiment}
Nanoscopium is an undulator beamline which, with a distance of about 150 m between the undulator and the experimental hutch, has been developed specifically for applications requiring excellent spatial coherence (\cite{Somogyi2015}). In our experiment monochromatic X-rays (starting at 7.1 keV) entered the experimental hutch. The coherent part of the beam was selected by an aperture, consisting of a pair of slits set to $1\times1$ mm$^2$. After, the X-ray beam passes through a tungsten pinhole of 24 \si{\micro\meter} diameter and hits the sample. The "Siemens star" gold test pattern consisted of 18 spokes fabricated by e-beam lithography and gold electroplating on a 500 nm thick Si3N4 membrane. The detector used to record the diffracted signal was an indirect imaging device based on a 70 \si{\micro\meter} thick Luag:Ce scintillator coupled with a scintillator to magnify the image onto a large fast and sensitive CMOS camera (Hamamatsu ORCA Flash 4.0 sCMOS). It was placed at 331 cm from the sample. The field of view covers 1.3 mm side and the measured spatial resolution is about 2 \si{\micro\meter} (FWHM). A microscope objective of $\times5$ magnification was used for preliminary alignment, allowing to locate the sample under the beam.

The X-ray intensity was first monitored using a 8 \si{\micro\m} thick Si diode, the X-ray energy was measured through the X-ray fluorescence spectra acquired by a silicon drift detector. Diffraction patterns were acquired using high-dynamic range: each pattern consisted of a long (10 s) and a short (1 s) acquisition. The long acquisition contains information at high scattering angles, the information at the saturated center of the CCD is obtained from the short acquisition (with a multiplicative factor of 10).
As the X-ray energy needed to be scanned over a long range the undulator separation and monochromator position needed to be varied. Special attention was payed to the stability of the undulator and monochromator during the scan to avoid artefacts in the broadband diffraction pattern due to a possible shift of the beam on the sample.

The phase retrieval of the synchrotron data was performed using the python library \textit{PyNX} \cite{Mandula2016, pynxsite}. The initial support was chosen to be a circle and the convergence was achieved after 32 iterations of a sequence of 20 error reduction (ER) \cite{Gerchberg1972} and 50 hybrid input-output (HIO) \cite{Fienup1978} algorithms. The images shown in Fig. \ref{fig:soleil_x4}c) and d) are the raw reconstructions, meaning that none post-treatment was performed on them. In this way, monochromatic and monochromatised reconstructions are comparable.

\end{document}

\title{Broadband Coherent Diffraction for Single-Shot Attosecond Imaging - Supplementary Information}

\author{Julius Huijts}
 \affiliation{LIDYL, CEA, CNRS, Université Paris-Saclay, CEA Saclay, Gif-sur-Yvette, France}
 
\author{Sara Fernandez}
\affiliation{LIDYL, CEA, CNRS, Université Paris-Saclay, CEA Saclay, Gif-sur-Yvette, France}

\author{David Gauthier}
\affiliation{LIDYL, CEA, CNRS, Université Paris-Saclay, CEA Saclay, Gif-sur-Yvette, France}

\author{Maria Kholodtsova}
\affiliation{LIDYL, CEA, CNRS, Université Paris-Saclay, CEA Saclay, Gif-sur-Yvette, France}

\author{Ahmed Maghraoui}
\affiliation{LIDYL, CEA, CNRS, Université Paris-Saclay, CEA Saclay, Gif-sur-Yvette, France}

\author{Kadda Medjoubi}
\affiliation{Synchrotron Soleil, BP 48, Saint Aubin, Gif-sur-Yvette, France}

\author{Andrea Somogyi}
\affiliation{Synchrotron Soleil, BP 48, Saint Aubin, Gif-sur-Yvette, France}

\author{Willem Boutu}
\affiliation{LIDYL, CEA, CNRS, Université Paris-Saclay, CEA Saclay, Gif-sur-Yvette, France}

\author{Hamed Merdji}\thanks{Corresponding Author: hamed.merdji@cea.fr}
\affiliation{LIDYL, CEA, CNRS, Université Paris-Saclay, CEA Saclay, Gif-sur-Yvette, France}

\date{\today}

\maketitle

\subsection{Movie 1}

Movie of the monochromatisation process of the broadband diffraction pattern from the experiment in the visible (main text, figure 2). As mentioned in the Methods section of the main text, the monochromatisation is performed in a regularised way by adding Krylov basis vectors. By increasing the number of basis vectors k, the matrix-vector problem is inverted, approaching the exact solution. In this video, the diffraction pattern is monochromatised by using up to k=20 basis vectors.

\subsection{Movie 2}
Movie of the monochromatisation process of the broadband diffraction pattern from the hard X-ray simulation at 20 \% bandwidth (Supplementary Information, figure S1), illustrating the behaviour of semi-convergence. As mentioned in the Methods section of the main text, the monochromatisation is performed by adding Krylov basis vectors. As the number of basis vectors k is increased, first the signal is inverted (up to about k=10), then gradually the inverted noise starts to dominate (up to k=60).

\subsection{Investigating the robustness at larger bandwidths - An applied case}
To investigate the robustness of our method at increasing bandwidth, we have simulated a broadband CDI experiment for various bandwidths. As an illustration of the applicability of our method, an EUV lithography mask was chosen as a sample.
EUV lithography masks consist of a patterned absorber (typically TaBN) deposited on a multilayer mirror on a quartz substrate. Provided the substrate is sufficiently thin and homogeneous, the absorber pattern can be imaged in transmission geometry through coherent diffractive imaging using hard X-rays, greatly simplifying the experimental setup while offering single nanometer resolution and being sensitive to defects in the multilayer mirror.

\begin{figure*}[t]
\includegraphics[width=1.0\textwidth]{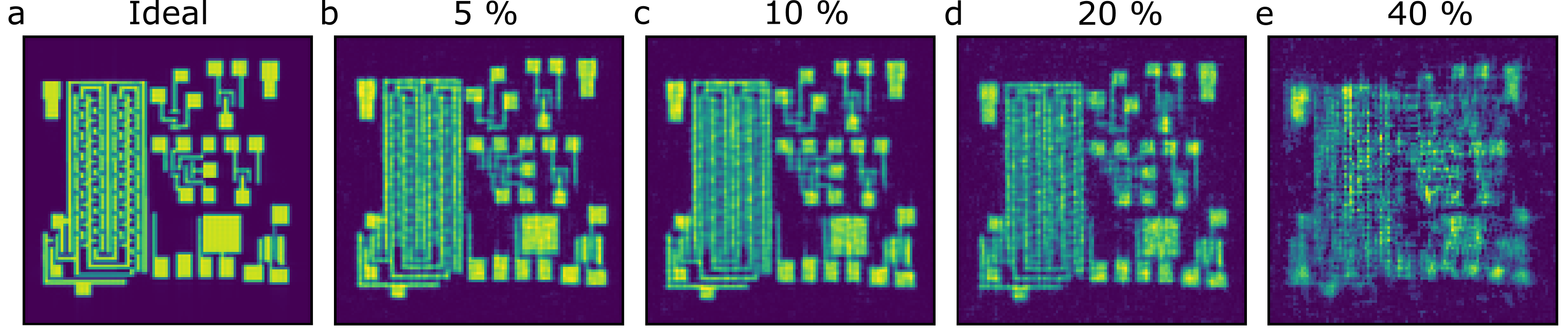}
\caption{\label{fig:xraysimu}The ideal reconstruction (a) and reconstructions for 5 to 40\% bandwidth (b-e) at a signal level of $10^{14}$ photons in the broadband pattern.}
\end{figure*}

Hence we simulated X-ray diffraction from a 60 nm thick TaBN pattern at different bandwidths around 8 keV, at different signal levels. Figure \ref{fig:xraysimu}
 shows CDI reconstructions of the absorber pattern for 5, 10, 20 and 40\% bandwidth at a signal level of $10^{14}$ photons per broadband pattern, next to an ideal reconstruction (the inverse Fourier Transform at 8 keV without noise). The relevant parameters of this simulated experiment are summarized in Table \ref{tab:simuvals}.
\begin{table}
  \centering
  \begin{tabular}{l c c}
	Variable              & Value & Unit \\ 
	\hline
	X-ray central energy & 8 & keV \\
	total number of photons & $10^{14}$ & photons \\
	sample dx              & 10 & nm \\
	mask thickness              & 60 & nm \\
	distance sample-detector              & 2 (2.5) & m \\
	detector size              & $512\times512$ & pixels \\
	detector pixel size & 30 & $\si{\micro\meter}$ \\
	reconstructed pixel size & 20 (25) & nm \\
\end{tabular}
  \caption[Values used for the broadband X-ray CDI simulations.]{Values used for the broadband X-ray CDI simulations. For bandwidths of 20 and 40\% the sample-detector distance was increased to achieve reconstruction despite these large bandwidths, at the expense of resolution (values in parentheses).}
  \label{tab:simuvals}
\end{table}
Where conventional CDI fails to reconstruct such complex shapes at bandwidths well below 5\%, through our method even the 20\% bandwidth case is reasonably reconstructed. The increased bandwidth (and thus a decrease in signal per energy slice) manifests itself through a gradual decrease in resolution and contrast, finally leading to a failed reconstruction at 40\% bandwidth. It is worth noting that initially no phase retrieval was achieved at the bandwidths of 20\% and 40\%, but sacrificing resolution by increasing the sample-detector distance (from 2.0 to 2.5 m) allows for phase retrieval even at these large bandwidths.

For the X-ray simulation use was made of \textit{Condor} \cite{Hantke2016}, a simulation tool for (flash) X-ray imaging. The sample “particle” was introduced as a 3D refractive index map (with a voxel size of 10 nm) after which \textit{Condor} was called to calculate the diffraction patterns while scanning the X-ray energy. The patterns were then summed to obtain the broadband diffraction pattern, after which they were monochromatized and used as input for the phase retrieval algorithm. 

\FloatBarrier
\bibliography{broadband}